\documentclass[twocolumn,preprintnumbers,amsmath,amssymb]{revtex4}
\usepackage{graphicx}
\usepackage{dcolumn}
\usepackage{bm}

\begin{document}
\title{Asymmetry of localised states in a single quantum ring: polarization dependence of excitons and biexcitons}
\author{H. D. Kim$^{1}$}
\author{K. Kyhm$^{2,3}$}\email{kskyhm@pusan.ac.kr}
\author{R. A. Taylor$^{1}$}\email{r.taylor1@physics.ox.ac.uk}
\author{G. Nogues$^{2}$}
\author{K. C. Je$^{4}$}
\author{E. H. Lee$^{5}$}
\author{J. D. Song$^{5}$}
\affiliation{$^{1}$Clarendon Laboratory, Department of Physics,
University of Oxford, Oxford, OX1 3PU, U.K} \affiliation{
$^{2}$Department of NANOscience, Institut N\'{e}el, CNRS, rue des
Martyrs 38054, Grenoble, France}\affiliation{$^{3}$Department of
Physics Education, RCDAMP, Pusan Nat'l University, Busan 609-735,
South Korea}\affiliation{$^{4}$College of Liberal Arts and
Sciences, Anyang University, Gyeonggi-do 430-714, South
Korea}\affiliation{$^{5}$Nano-Photonics Research Center, KIST,
Seoul, 136-791, South Korea}
\date{\today}
\begin{abstract}
We performed spectroscopic studies of a single GaAs quantum ring
with an anisotropy in the rim height. The presence of an
asymmetric localised state was suggested by the adiabatic
potential. The asymmetry was investigated in terms of the
polarization dependence of excitons and biexcitons, where a large
energy difference ($\sim0.8\,$meV) in the exciton emission energy
for perpendicular polarizations was observed and the oscillator
strengths were also compared using the photoluminescence decay
rate. For perpendicular polarizations the biexciton exhibits twice the
energy difference seen for the exciton, a fact that may be attributed to a possible change in the
selection rules for the lowered symmetry.
\end{abstract}
\maketitle

Currently, quantum ring (QR) structures are of
great interest for the optical Aharonov-Bohm (AB) effect
 \cite{Bayer,Govorov1,Govorov2,Holland1,Holland2}. While the
rotating charge in the shell of a type-II quantum dot (QD)
determines the AB oscillation period, the orbital radius
difference of the electrons and holes is the crucial parameter in
a QR, for which the coupling to the magnetic flux is of opposite
sign. Nevertheless, the individual behaviour of each of the
particles is not clear in a QR; either the radius is larger or one
of them is localised as in the case of a type-II QD.

Recent measurements have shown that the morphology of a QR
is anisotropic, where the rim height is not constant around the
azimuthal angle. In the case of this so-called volcano-like
structure \cite{Holland2}, the azimuthal quantum number is no
longer valid, and the wavefunction in the anisotropic QR can be
either localised or delocalised through tunnelling \cite{Taiwan}.
Nevertheless, a persistent current can be observed in a QR as
evidence of AB-oscillation, when an external magnetic field is
large enough to overcome the anisotropy-induced potential barrier
or asymmetric exchange interaction \cite{Holland1}. Asymmetry and
anisotropy seem to have been overlooked in the spectroscopy of QRs
\cite{Bayer,Govorov1,Govorov2,Kuroda,Warburton,Spain}. In this
work, the presence of a localised state in a single
GaAs/Al$_{0.3}$Ga$_{0.7}$As QR arises from the volcano-like QR
structure, which corresponds to an excited state of the vertical
confinement. Also, the asymmetry of the localised state has been
investigated in terms of the polarization dependence of excitons
and biexcitons.

GaAs rings \cite{Kuroda} were grown on an n-doped GaAs
(001) substrate using a molecular beam epitaxy system with an ion
getter pump. After thermal cleaning of the substrate under arsenic
ambient at 600$^{\circ}$C, a 100 nm-thick GaAs buffer layer and a
50 nm-thick Al$_{0.3}$Ga$_{0.7}$As layer were grown successively
at 580$^{\circ}$C. The substrate temperature was decreased to
310$^{\circ}$C, and Ga metal equivalent to 2 monolayers of GaAs
was introduced to the substrate at the main chamber pressure of
$\sim3\times10^{-10}\,$Torr. When the substrate temperature
reached 200$^{\circ}$C, arsenic tetramers were introduced to form
GaAs rings. Finally, the rings were capped with 60 nm-thick
Al$_{0.3}$Ga$_{0.7}$As and 3 nm-thick GaAs for optical
measurements. The photoluminescence (PL) of a single QR was
collected at $4\,$K using a confocal arrangement, where
frequency-doubled ($400\,$nm) Ti:sapphire laser pulses ($120\,$fs
pulse duration at a $80\,$-MHz repetition rate) were focused on
the QR sample ($\sim6\,$QRs/$\mu$m$^{2}$) with a spot-size of
$0.8\,\mu$m$^{2}$. A time-correlated single photon counting system
was used to obtain the time-resolved PL (TRPL).\\
\begin{figure}
\includegraphics[width=9 cm]{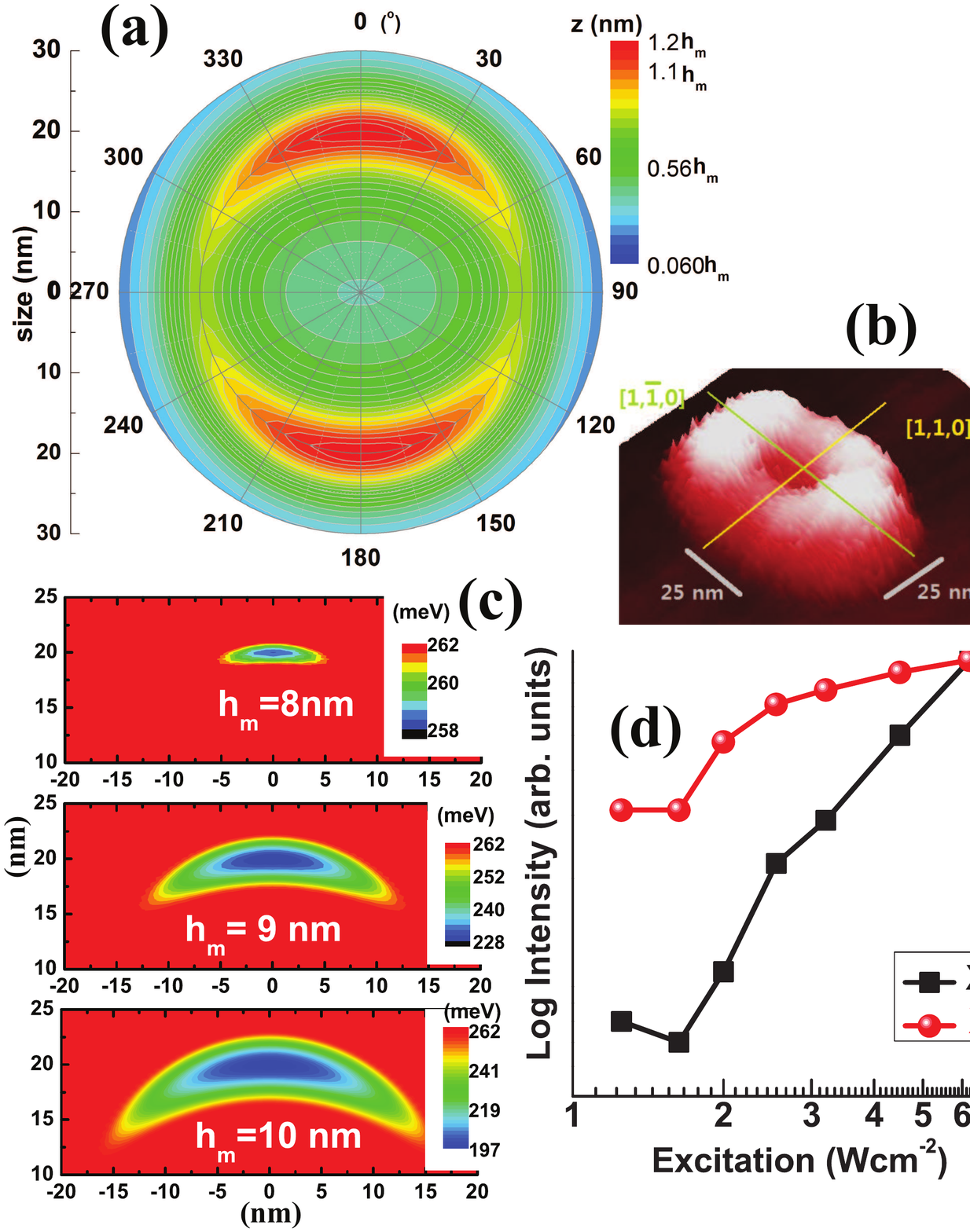}
\caption{\label{fig:pl} (a) Volcano-like ring model for the rim
height based on the AFM morphology (b), where the localised
adiabatic potentials of the electron for a vertical quantum number
of $k=3$ are shown for three rim height parameters
($h_{\rm{m}}=8,9,10\,$nm) (c). (d) PL intensity corresponding to
the $k=3$ states from the XX is compared with that from the X with
increasing the excitation power.}
\end{figure}
As shown in Fig.1(a), the volcano-like ring structure was
modelled in cylindrical coordinates ($z=z(r,\phi)$) based on
atomic force microscope (AFM) images of uncapped GaAs QRs
(Fig.1(b)), where both anisotropy and asymmetry are present
\cite{parameters,Kuroda}. The rim height is maximum at the
azimuthal angles of $0^{\circ}$ and $180^{\circ}$ along the
$[1\bar{1}0]$ direction, and a minimum at the perpendicular angles
of $90^{\circ}$ and $270^{\circ}$ along the $[110]$ direction,
respectively. The in-plane shape is elliptical with the long axis
along $[1\bar{1}0]$, and the height in the middle of the QR is
$\sim3\,$nm. Since the vertical height ($7\sim12\,$nm) is smaller
than the ring size ($\sim50\,$nm), the fast vertical wavefunction
can be separated by the adiabatic approximation \cite{Holland2},
where the potential of electron and hole depends on the
volcano-like ring structure respectively as
$V_{\rm{e,h}}(z,r,\phi)\simeq V_{\rm{e,h}}(z(r,\phi))$.
Consequently, the adiabatic potential
$\varepsilon^{k}_{\rm{e,h}}(r,\phi)$ can be obtained for electron
and hole separately by solving the vertical part of the
Schr$\ddot{\rm{o}}$dinger equation, where the vertical confinement
is represented by the vertical quantum number $k$. In the case of
an ideal isotropic ring, $\varepsilon^{k}_{\rm{e,h}}(r)$ can be
simplified by using a parabolic function ($\sim(r-r_{0})^{2}$),
where the maximum ring height is positioned at $r_{0}$
\cite{Song}. Although the fine structure of the PL spectrum in a
QR  has often been attributed to quantized rotational motion along
the rim \cite{Kuroda}, the cylindrical symmetry of a QR can easily
break down, which is similar to the case of an elliptical QD.
Also, the anisotropy in the rim height requires an azimuthal
angle-dependence of the vertical confinement. Both the asymmetry
of the in-plane ellipticity and the height anisotropy have been
mostly overlooked in previous spectroscopy. These effects can be
paramaterized in terms of $\varepsilon^{k}_{\rm{e,h}}(r,\phi)$,
which is similar to an inversion of the asymmetric and anisotropic
structure morphology.

\begin{figure}
\includegraphics[width=8 cm]{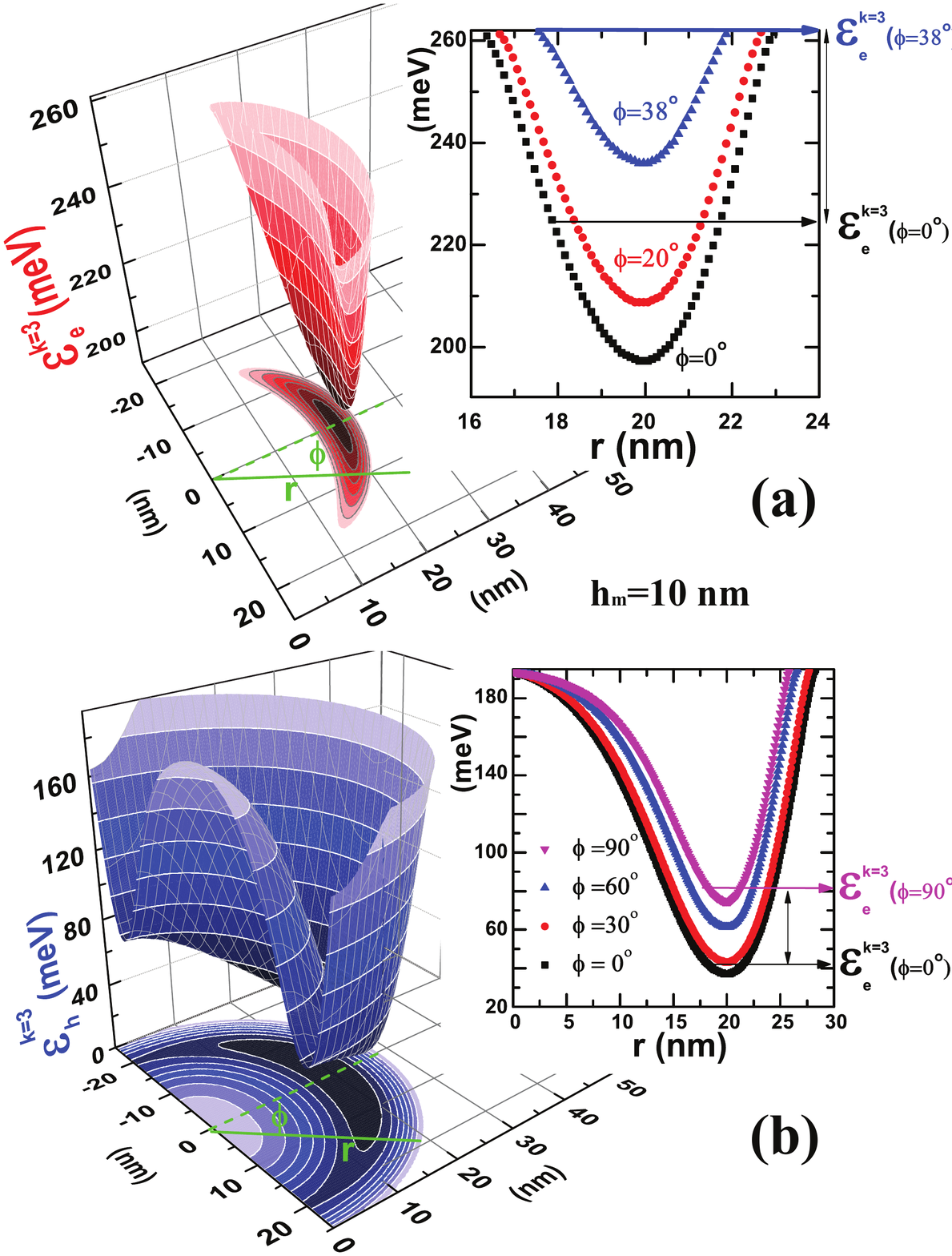}
\caption{\label{fig:p2} Adiabatic potentials
($\varepsilon^{k=3}_{\rm{e,h}}$) of the electron (a) and the hole
(b) for a vertical quantum number of $k=3$ are shown, where the
insets show the energy range of the radially confined levels for
the azimuthal angle for the electron and the hole, respectively.}
\end{figure}
Provided that the potential hill in
$\varepsilon^{k}_{\rm{e,h}}(r,\phi)$ is found near the azimuthal
angles $90^{\circ}$ and $270^{\circ}$, the wavefunction
delocalisation of the ground state depends on the tunneling
efficiency, which determines either localised \cite{Taiwan} or
extended states. However, it should be remembered that the number
of confinement states can be a measure of the confinement size,
i.e., the higher rim ($z$) contains more vertical confinement
states than the lower rim. In other words, excited confinement
states are allowed only at the limited azimuthal angles of the
high rim region as the $k$ quantum number increases
\cite{Holland2}. We found that vertical confinement states can be
defined at all of the azimuthal angles up to $k=2$. Therefore, the
vertical excited state of $k=3$ is a criterion of the
localisation. As shown in Fig.1(c), the electron adiabatic
potentials for $k=3$ ($\varepsilon^{k=3}_{\rm{e}}(r,\phi)$) are
localised at limited azimuthal angles for three different QR
heights ($h_{\rm{m}}=8,9,10\,$nm), where the rim height is
characterized by a parameter $h_{\rm{m}}$\cite{parameters}. As
the QR height is decreased, the localized potential area becomes
reduced.

The adiabatic potentials of the electron
($\varepsilon^{k=3}_{\rm{e}}$) and hole
($\varepsilon^{k=3}_{\rm{h}}$) are compared for
$h_{\rm{m}}=10\,$nm in Fig.2 \cite{comment}. Also, the radially
confined energy of the adiabatic potential for $\phi=0^{\circ}$
and $\phi=90^{\circ}$ are estimated with the parabolic
approximation for the electron and the hole (insets in Fig.2),
respectively. Whilst the holes are confined for all azimuthal
angles (Fig.2-(b)), the electron is confined to the limited
azimuthal angles ($-38^{\circ}\sim38^{\circ}$) (Fig.2-(a)) due to
the shallow potential well when compared to the barrier energy in the
conduction band ($262\,$meV). Therefore, the ground eigen-energy
of the electron and the hole should be located in the range of the
radially confined levels for allowed azimuthal angles
($\varepsilon^{k=3}_{\rm{e,h}}(\phi)$). For example, suppose the
ground state of the electron is located near $240\,$meV, then the
confined wavefunction becomes localised in a crescent-like structure.
However, the tunnelling wavefunction is extended up to
$\sim4.5\,$nm in the barrier. On the other hand, the hole
wavefunction also becomes localised when the hole ground state is
assumed to be $\sim60\,$meV, but the tunnelling length is small
($\sim0.7\,$nm) as the barrier energy in the valence band
($195\,$meV) is still large. Consequently, the localised area of
the electron would be larger due to the large tunnelling, and PL
quenching becomes significant with increasing temperature. We found
the PL intensity is nearly quenched beyond $\sim45\,$K.

The exact ground state energy of the electron-hole (e-h)
pair can be refined by adding the Coulomb interaction to the
independent e-h pair, i.e., the total energy is determined by the
ground state energy sum of each adiabatic potential and the
Coulomb energy. In the case of strong confinement, the effective
Coulomb interaction is known to be enhanced by a factor of
1.786. However, both the vertical and the lateral confinement
energy are far larger than the Coulomb interaction ($\sim4.2\,$meV
for bulk GaAs). Thus, the energy levels are dominated by the
confinement effect, and this rough model predicts the energy range of the
e-h pair at the $k=3$ state ($1.814\sim1.864\,$eV). As shown in
Fig.3, the PL spectrum was measured near the barrier
(Al$_{0.3}$Ga$_{0.7}$As) energy. It exhibits a strong polarization
dependence for the exciton (X) and biexciton (XX) states at
$1.6\,$kWcm$^{-2}$, where the nature of the XX emission appearing
$5\,$meV below that of the X was characterised by a quadratic rise
in its PL intensity (Fig.1(d)) and the fast decay of the TR-PL
(Fig.4(b)) relative to the X. The X and XX emission spectra were
both measured simultaneously at a series of analyzer angles. We
also observed the PL spectrum in the predicted $k=3$ range
($1.813\,$eV, $1.821\,$eV, $1.832\,$ eV, and $1.842\,$eV) at
different QRs; they all show the similar polarization
dependence.

\begin{figure}[t]
\includegraphics[width=8 cm]{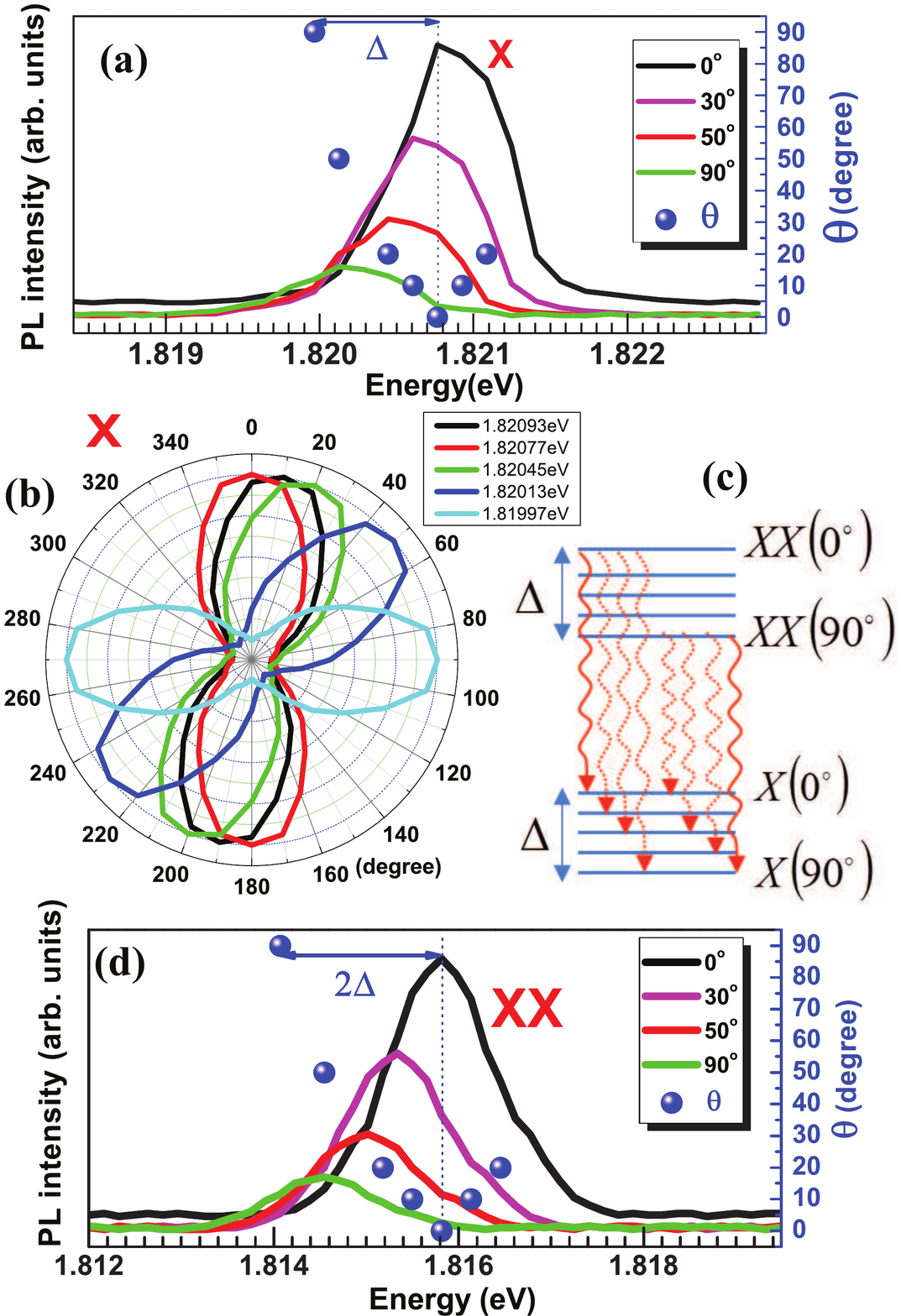}
\caption{\label{fig:p3} PL spectrum of the X (a) and XX (d) at
various analyzer angles in a single QR. (b) Normalized PL
intensity as a function of analyzer angle is mapped in polar
coordinates at various energies, whereby the relative angular
delay ($\theta$) of each dumbbell-like trace can be obtained for
the X and XX. (c) Schematic diagram of a multitude of transitions
between the fine structure states of XX and X.}
\end{figure}
The wavefunction of the electron and the hole can be
imagined roughly as an inversion structure of the adiabatic
potential in Fig.2. Since both $\varepsilon^{k=3}_{\rm{e}}$ and
$\varepsilon^{k=3}_{\rm{h}}$ are anisotropic, the e-h pair is
likely to be localised in either of the two crescent structures
instead of delocalization around the whole rim. The localised e-h
pair can be verified in terms of the large energy splitting for
perpendicular polarizations. Since the localised states are of a
crescent shape, the fine structure states resulting from various
different confinement dimensions  may depend on the analyzer
angle. Additionally, during sample growth, identical crescents
are unlikely to form in an anisotropic QR and the 
resonance at the $k=3$ state between the two crescents is vulnerable
to small size differences. In this case, localisation of the e-h pair is favored 
at the larger crescent-like structure.

Whilst the two orthogonally polarized states of
$|X\rangle$ and $|Y\rangle$ in an elliptical QD can be spectrally
isolated by a linear polarizer \cite{Grenoble,HtoonPRL,HtoonPRB},
the fine structure states at different angles in the asymmetric
$\varepsilon^{k=3}(r,\phi)$ can overlap in the PL spectrum of X.
Therefore, the analyzer angle dependence of the PL intensity is
mapped at different energies in polar coordinates as shown in
Fig.3(b), where the maximum intensity of each PL spectrum is
normalized to make a comparison possible. For example, the maximum
PL intensity of X at $1.8208\,$eV measured at $0^{\circ}$ is
gradually reduced until the analyzer angle rotates to
$90^{\circ}$, but increases again up to $180^{\circ}$.
Consequently, this sinusoidal behavior gives rise to a
dumbbell-like trace. All of the dumbbell-like traces obtained at
different energies are similar, but rotated in a regular manner.
The relative angular delay ($\theta$) between the traces varies
with emission energy. The trace at $1.8200\,$eV is
$90^{\circ}$-rotated with respect to that at $1.8208\,$eV, and an
energy difference ($\Delta$) of $0.8\,$meV is obtained. This value
is remarkably large in comparison to the asymmetric splitting
($\Delta_{XY}\sim0.1\,$meV) of an elliptical quantum dot.\\
\indent In the case of an elliptical quantum dot, an asymmetric
electron-hole exchange interaction leads to a splitting
($\Delta_{XY}$) of the double exciton state ($|\pm1\rangle$) into
two singlet states
($|X,Y\rangle=(|+1\rangle\pm|-1\rangle)/\sqrt{2}$), where two
linearly and orthogonally polarized dipoles ($|X\rangle$ and
$|Y\rangle$) are defined along the principal axes of an elliptical
QD \cite{Grenoble,HtoonPRL,HtoonPRB}. This also gives rise to the
same splitting of the biexciton emission ($\Delta_{XY}$), which
involves the transition from biexciton to two singlet exciton
states ($|X,Y\rangle$), respectively. However, the energy
difference for the perpendicular polarization in the XX spectrum
(Fig.3(d)) is nearly twice ($\sim1.75\,$meV) that of X, where the
angular delay of $90^{\circ}$ is measured by the dumbbell-like
trace of XX. We have also confirmed this phenomenon in QRs of
different sizes.

When considering XX in this structure, two kinds of
generation are possible; either two Xs in the same localised
crescent structure or separate Xs located in two identical
crescent structures are bound, respectively. Although the binding
energy of the latter case was known to be sub-meV\cite{Taiwan},
the generation can be inhibited by the small size difference
between the two crescent structures. For large GaAs QDs
($30\sim40\,$nm in radius), the XX binding energy is known to be a
few meV \cite{Ikezawa,GaAschina}. Therefore, the large binding
energy ($\sim5\,$meV) in this result is likely to arise from a
localised XX, which consists of two Xs localised at the same
crescent structure. The large asymmetric splitting
($\Delta_{XY}\sim1.75\,$meV) of the XX also supports a localised
XX. When the symmetry is lowered, it is known that the selection
rules change significantly \cite{Deformation,Woggon}. This can
give rise to an increase in the number of dipole-allowed
transitions, i.e., a multitude of transitions between the
two-exciton and one-exciton states, resulting in a broad PL
spectrum for the XX. It is also noticeable that the XX PL spectrum
in Fig.3(d) is rather broad. Consequently, as shown schematically
in Fig.3(c), a multitude of transitions between the fine structure
states of XX and X, which are denoted by the relative angular
delay ($\theta$), may result in twice the splitting when compared
to the X for perpendicular polarizations because of the change in
the ideal selection rules.

\begin{figure}[t]
\includegraphics[width=8cm]{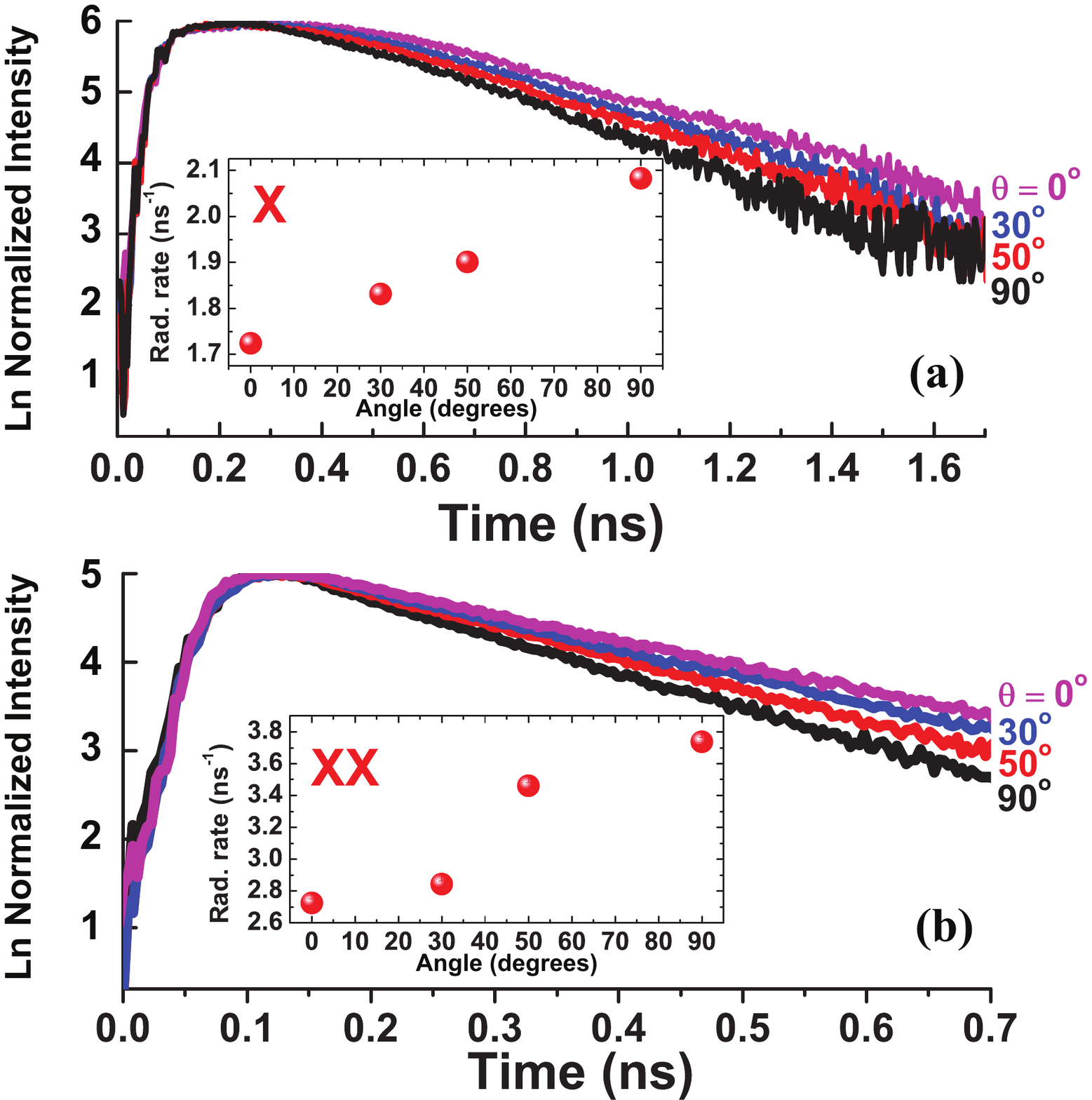}
\caption{\label{fig:p4} TRPL of the of the localised state of X
(a) and XX (b) in a single QR measured at $1.6\,$kWcm$^{-2}$ for
various analyzer angles, where each PL rate shown in the inset.}
\end{figure}
Although the wavefunction distribution of the fine
structure states is not clear, the oscillator strength difference
of these states can be observed in terms of the size dependence of
the PL decay rate, where the PL spectra for different confinement
sizes are isolated by the analyzer angle. As shown in Fig.4, TRPL
of X and XX was measured at various analyzer angles ($\theta$),
where the PL decay rates were obtained by a linear fit to the
monotonic decay section on a log scale (shown in the inset). We
found that the PL decay rates of both X and XX increase for
increasing analyzer angle up to $90^{\circ}$. Interestingly, XX
shows a novel feature in the size dependent oscillator strength.

Whilst the radiative recombination of the e-h pair is
characterized by a single exponential decay after $\sim400\,$ps, a
plateau range is observed up to $\sim300\,$ps in Fig.4(a). This
may be associated with extended states of the electron. Initially,
e-h recombination occurs at the wavefunction overlap range
between the localised electron and hole. However, extended states
keep feeding electrons to localised states. State
saturation in the plateau suggests the presence of a feeding
source for intra-relaxation of  localised electrons
($k=3$) to lower states ($k=2$ or $k=1$). Furthermore, it is
noticeable that the decay time ($532\pm56\,$ps) of the X is
longer than that seen in GaAs QDs ($250\pm50\,$ps)
\cite{Grenoble}. When compared to the recombination rate of a hole
near localised electrons, the recombination rate of a hole far
from such localised electrons can be reduced. This effect possibly
results in the reduction of the total decay rate in the $k=3 $ localised structure.

In conclusion, the presence of an asymmetric localised
state in a single GaAs/Al$_{0.3}$Ga$_{0.7}$As QR was observed by
its polarization dependence in the PL spectrum, which is in
agreement with the adiabatic potential arising from a volcano-like
anisotropic morphology. The fine structure states of the crescent-like adiabatic
potential were resolved by changing the analyzer angle, whereby a
large energy difference in the exciton ($\sim0.8\,$meV) and
biexciton ($\sim1.75\,$meV) splitting was observed for
perpendicular polarizations, and oscillator strength
differences were also compared in terms of the PL decay rate.

\begin{acknowledgments}
This work was supported by the National Research Foundation of
Korea Grant funded by MEST (NRF-2011-013-C00025 and
NRF-2010-008942) and GRL program, and the KIST institutional
program. Authors are also grateful to the late Dr. Lee for
organizing the international collaboration.
\end{acknowledgments}


\begin{thebibliography}{}
\bibitem{parameters}%
 $R=20\,$nm, $h_{0}=4\,$nm, $h_{\infty}=0.4\,$nm,
 $\gamma_{0}=5.5\,$nm, $\gamma_{\infty}=5.5\,$nm, $\xi_{\rm{h}}=0.2$, $\xi_{\gamma}=0$, and
 $\xi_{\rm{R}}=0$, where all parameters are defined in
 \cite{Holland2}, and the effective mass and band-offset of electron
 and hole of GaAs and AlGaAs are given in \cite{Kuroda}.
\bibitem{comment}%
While $k=3$ is the largest vertical quantum number for electron,
higher $k(>3)$ are possible in the valence band. However, the
corresponding energy is beyond the observed PL spectrum, and the
accuracy is limited for ignorance of the spin-orbit coupling.
\end{thebibliography}
\end{document}